 \documentclass[aps,prc,preprint,groupedaddress,floatfix]{revtex4-2}

\usepackage{amsmath}

\usepackage{graphicx}

\def\m@thcombine#1#2{%
  \setbox0=\hbox{$#1$}
  \setbox1=\hbox{$#2$}
  \ifdim\wd0>\wd1
    \setbox0=\hbox to\wd1{\hss\box0\hss}
  \else
    \setbox1=\hbox to\wd0{\hss\box1\hss}
  \fi
  \mathop{\vcenter{
    \offinterlineskip\box0\box1}}}
\def\lesim{\m@thcombine<\sim}
\def\gesim{\m@thcombine>\sim}
\def\lessgtr{\m@thcombine<>}
\def\gtrless{\m@thcombine><}
\newcommand{\bra}[1]{\left\langle #1 \right|}
\newcommand{\ket}[1]{\left| #1 \right\rangle}
\newcommand{\braket}[3]{\left\langle #1 | #2 | #3 \right\rangle}

\newcommand{\vecr}{\mbox{\boldmath $r$}}
\newcommand{\vecrs}{\mbox{\boldmath $r$}\sigma}
\newcommand{\vecrst}{\mbox{\boldmath $r$}\tilde{\sigma}}
\newcommand{\vecrsp}{\mbox{\boldmath $r$}'\sigma'}

\newcommand{\rhot}{\tilde{\rho}}

\newcommand{\calG}{{\cal{G}}}

\newcommand{\calA}{{\cal{A}}}
\newcommand{\calB}{{\cal{B}}}

\newcommand{\del}{\partial}
\newcommand{\eps}{\epsilon}

\newcommand{\Phat}{\hat{P}}

\newcommand{\beq}{\begin{equation}}
\newcommand{\eeq}{\end{equation}}
\newcommand{\bal}{\begin{align}}
\newcommand{\eal}{\end{align}}

\newcommand{\up}{\uparrow}
\newcommand{\down}{\downarrow}

\newcommand{\Ecal}{\mathcal{E}}

\begin{document}

\title{
Higgs response and pair condensation energy in superfluid nuclei}

\author{Kengo Takahashi}
\author{Yusuke Matsuda}
 \affiliation{Graduate School of Science and Technology, Niigata University, Niigata 950-2181, Japan}
\author{Masayuki Matsuo}
\affiliation{Department of Physics, Faculty of Science, Niigata University, Niigata 950-2181, Japan}
\email{matsuo@phys.sc.niigata-u.ac.jp}

\date{\today}

\begin{abstract}
The pairing correlation in nuclei causes a characteristic excitation, known as 
the pair vibration, which is populated by the pair transfer reactions.
 Here we introduce a new method of characterizing the pair vibration by employing an analogy
 to the Higgs mode, which emerges in  infinite superconducting/superfluid systems as a
 collective vibrational mode associated with the amplitude oscillation of the Cooper pair condensate.
The idea is formulated by defining a pair-transfer probe, the Higgs operator,
and then describing the linear response and the strength function to this probe.  We will
show that the pair condensation energy in nuclei can be extracted with use of
 the strength sum and the static polarizability of the Higgs response.
 In order to demonstrate and validate the method, we perform for Sn isotopes 
 numerical analysis based the quasi-particle random phase approximation to the 
 Skyrme-Hartree-Fock-Bogoliubov model.
We discuss a possibility to apply this new scheme to pair transfer experiment.
\end{abstract}

\maketitle

\section{Introduction}\label{sec:intro}
Pair correlation is one of the most important many-body correlations influencing the ground state
structure and low-energy excitations in nuclei\cite{Brink-Broglia,Broglia-Zelevinsky,Dean-Hjorth-Jensen,BM2,Ring-Schuck}.
It also brings about superfluidity or superconductivity
in infinite nuclear matter and in neutron stars\cite{Broglia-Zelevinsky,Dean-Hjorth-Jensen,Takatsuka-Tamagaki,Sedrakian-Clark}.
 It originates from 
attractive interactions between
like nucleons (or unlike nucleons), depending on the quantum numbers of the
interacting pair, eg $^{1}S_0$, $^{3}S_1$, $^{3}P_2$, etc. It causes condensation of
the correlated pairs, Cooper pairs. 

One of the most important aspects of the superfluidity/superconductivity in Fermion systems
is the presence of the pair gap $\Delta$, which originates from the 
condensation of the Cooper pairs\cite{BCS}. The pair gap or the pair condensate is the 
order parameter of the superfluidity/superconductivity and it plays also a role of the dynamical variable in the 
phenomenological Ginzburg-Landau theory\cite{Ginzburg-Landau}. Another important feature is
that the finite pair gap breaks the U(1) gauge symmetry. The spontaneous symmetry breaking
produces characteristic collective modes of excitations, known as the 
Anderson-Boboliubov (Nambu-Goldstone) mode\cite{Anderson1958,Bogoliubov1958,Nambu1960} and 
the Higgs mode\cite{Anderson1958,Higgs1964,Littlewood-Varma1981-82,Pekker-Varma,Shimano-Tsuji}, which are
oscillations of the order parameter with respect to the phase and the amplitude,
respectively. The Anderson-Bogoliubov mode is a mass-less phonon with a linear dispersion relation
whereas the Higgs mode is massive, i.e. the lowest excitation energy is $2\Delta$\cite{Anderson1958,Higgs1964,Littlewood-Varma1981-82,Pekker-Varma,Shimano-Tsuji}.

An analogy to  the superfluidity/superconductivity 
has been brought into finite nuclei 
by applying the BCS theory to the nuclear Hamiltonian\cite{Bohr-Mottelson-Pines,Belyaev1959,BM2,Ring-Schuck}. 
In this description, nuclei may be
in superconducting/superfluid or normal phase depending on whether they are closed
or open shell nuclei, and also on other conditions such as rotational frequency\cite{Brink-Broglia,Broglia-Zelevinsky}. 
The odd-even mass difference is a typical indicator of the nuclear superfluid phase\cite{BM2,Ring-Schuck}
since it corresponds to the pair gap in the single-particle excitation spectrum.
 Two-nucleon transfer (pair transfer) reaction has been considered as a probe for the pair correlation
in nuclei\cite{Yoshida62, Bes-Broglia66, Broglia73, Brink-Broglia, BM2, Oertzen2001,Potel2013}. 
A typical example is the $(p,t)$ and $(t,p)$ reaction on the even-even Sn isotopes,
which shows enhanced cross sections for
 transitions between 0$^{+}$ ground states of adjacent isotopes\cite{Yoshida62,Bes-Broglia66,Broglia73,Brink-Broglia}.
In parallel to 
the rotational motion of deformed nuclei, the concept  of the
pair rotation  is introduced\cite{Bes-Broglia66,Broglia73,Brink-Broglia}, as a nuclear counter part of the 
Anderson-Bogoliubov mode\cite{Anderson1958,Bogoliubov1958}.
It has been argued that the pair rotation is a phase rotation of the gap, and is
characterized with a rotational band,
a series of 0$^{+}$ ground states along the isotope chain. The 
transition strength of the pair transfer within the rotational band is
related to the order parameter, the pair gap. 

Another collective mode associated with the pair correlation is 
the pair vibration\cite{Bes-Broglia66,Broglia73,Brink-Broglia},
which is introduced in an analogy to the shape vibration modes in nuclei.
It is a collective vibrational state with spin parity 0$^{+}$ and excitation energy around
$\sim 2\Delta$. It is regarded as 
oscillation of the amplitude of the pair gap $\Delta$. 
The lowest-lying excited 0$^{+}$ states populated
by  the $(p,t)$ and $(t,p)$ reactions or other two-neutron transfer are identified as 
the pair vibrational states.  Although not stated explicitly in the preceding works, 
the low-lying pair vibrational mode might be regarded as a counter part  of the Higgs mode 
in superfluids/superconductors. 

We remark here that  microscopic theories such as the 
quasiparticle random phase approximation predict also another pair vibrational mode 
with high excitation energy, called often the giant pair vibration (GPV) \cite{Broglia1977,Cappuzzello2015,Cavallaro2019,Assie2019},
which arises from the shell structure inherent to finite nuclei.
It is therefore a non-trivial issue how the pair vibrations in nuclei, either low-lying or high-lying, 
or both, are related
 to the macroscopic picture of the Higgs mode. Recently an observation of
 the giant pair vibrations in a two-neutron transfer reaction populating $^{14}$C and $^{15}$C 
 has been reported\cite{Cappuzzello2015,Cavallaro2019}.  Identification in medium and heavy nuclei 
 is an open question\cite{Assie2019} and further experimental study is awaited.

In the present study, we intend to describe the pair vibrations, including both low-lying and
high-lying ones, from a view point of the Higgs mode. We shall discuss then that this viewpoint
 provides a new perspective to the nature of the pair correlation in nuclei.

Our study proceeds in two steps. Firstly, we 
introduce a new scheme of exploring a
response that probes the Higgs mode in a pair correlated nucleus.
One of commonly adopted methods to describe the pairing collective modes, including the
pair vibrations, is to consider addition or removal of a two-nucleon pair, or
describe transition strength for pair-addition and pair-removal operators. 
In the present work, we propose a variant of
 these operators, which is more suitable for probing the Higgs mode in finite nuclei. We formulate
 then the strength function for this new two-nucleon transfer operator, named Higgs
 strength function, and we then characterize the 
 pair vibrations, including both low-lying and high-lying, in terms of the Higgs strength function.
  The formulation is based on Skyrme-Hartree-Fock-Bogoliubov mean-field model and the
quasiparticle random phase approximation, which are widely used to describe the
ground state and collective excitations, including pair vibrations, 
in medium and heavy nuclei\cite{Ring-Schuck,Nakatsukasa-etal2016,Paar-etal2007}.  

Secondly, we shall argue that the Higgs strength function carries an important
information on the pair correlation in nuclei, in particular, the pair condensation energy or the effective
potential energy, which is the 
energy gain obtained by the condensation of Cooper pairs.  
Similarly to the Ginzburg-Landau theory of superconducting systems, an effective potential energy as a function
of the order parameter, i.e, the pair condensate, can be considered also in finite nuclei, and it
is straightforwardly evaluated in the framework of the Hartree-Fock-Bogoliubov mean-field model. 
On another hand,  the Higgs strength function probes the small amplitude oscillation of the order parameter
and one can expect that it can be related to  the curvature of the effective potential. Motivated by this idea,
we examine in detail the effective potential energy obtained with the present Skyrme-Hartree-Fock-Bogoliubov model. 
As shown later, the effective potential has a rather simple structure, parametrized with a fourth order
polynomials of the pair condensate. Combining these observations, we propose a novel method with which one can 
evaluate the pair condensation energy using the Higgs strength function. We shall discuss the validity of this
method with numerical calculations performed for neutron pairing in Sn isotopes.

 \section{Theoretical framework}
 
As a theoretical framework for our discussion, we adopt the 
Hartree-Fock-Bogoliubov (HFB) theory in which the pair correlation
is described by means of the Bogoliubov's quasiparticles and the 
selfconsistent pair field\cite{Ring-Schuck}. It has the same structure as that of 
the Kohn-Sham  density functional theory for superconducting electron systems 
with an extension based on the Bogoliubov-de Genne scheme\cite{Nakatsukasa-etal2016}.
The adopted model is essentially the same as the one employed in our previous studies\cite{Serizawa2009,Matsuo2010,Shimoyama2011,Shimoyama2013}.
We here describe an outline of the model with emphasis on some aspects which are
necessary in the following discussion.

In the HFB framework, the superfluid/superconducting state is described with a
generalized Slater determinant $\ket{\Psi}$, which is a determinantal state
of Bogoliubov's quasiparticles. The total energy of the system for $\ket{\Psi}$
is a functional of one-body densities of various types, and we adopt
the Skyrme functional $\Ecal_{Skyrme}$
combined with the pairing energy $\Ecal_{pair}$.
Assuming that the
pairing energy originates from a contact two-body interaction,  $\Ecal_{pair}[\rho(\vecr),\rhot(\vecr),\rhot^*(\vecr)]$
is an functional of the local pair condensate (also called pair density in the literature)
\begin{align}
\tilde{\rho}(\vecr) &= \braket{\Psi}{\hat{P}(\vecr)}{\Psi}
= \bra{\Psi} \psi(\vecr \up)\psi(\vecr \down)\ket{\Psi},
\ \ \ &
\hat{P}(\vecr)=\frac{1}{2}\sum_\sigma \psi(\vecrst)\psi(\vecrs), \\
\tilde{\rho}^*(\vecr) &= \braket{\Psi}{\hat{P}^\dagger(\vecr)}{\Psi}
= \bra{\Psi} \psi^\dagger(\vecr \down)\psi^\dagger(\vecr \up)\ket{\Psi},
\ \ \ &
\hat{P}^\dagger(\vecr)=\frac{1}{2}\sum_\sigma \psi^\dagger(\vecrs) \psi^\dagger(\vecrst),
\end{align}
and local one-body density
\begin{align}
\rho(\vecr) &= \braket{\Psi}{\hat{\rho}(\vecr)}{\Psi}, \ \ \ & \hat{\rho}(\vecr)=\sum_\sigma \psi^\dagger(\vecrs)\psi(\vecrs) 
\end{align}
where 
$\psi^\dagger(\vecrs)$ and $\psi(\vecrs)$ are nucleon creation and annihilation operators
with the coordinate and the spin variables. Note that the local pair condensate $\rhot(\vecr)$ and $\rhot^*(\vecr)$ 
is finite 
if the condensation of Cooper pairs occurs.  
Here and hereafter we do not write the isospin index  for simplicity. Notation follows Ref.\cite{Matsuo2001}.

\noindent
\underline{Stationary ground state}

The variational principle  $\delta \Ecal=\delta \Ecal_{Skyrme} + \delta \Ecal_{pair}=0$ leads to the
so-called Hartree-Fock-Bogoliubov equation or the Bogoliubov-de-Genne equation
\begin{equation}\label{HFB}
\left(
\begin{array}{cc}
\hat{t}+\Gamma(\vecr)-\lambda & \Delta(\vecr) \\
\Delta^*(\vecr) & -(\hat{t}+\Gamma(\vecr)-\lambda) 
\end{array}
\right)
\left(
\begin{array}{c}
\varphi_{1,i}(\vecrs) \\
\varphi_{2,i}(\vecrs) 
\end{array}
\right)
=E_i 
\left(
\begin{array}{c}
\varphi_{1,i}(\vecrs) \\
\varphi_{2,i}(\vecrs) 
\end{array}
\right),
\end{equation}
which determines energy $E_i$ and two-component wave function $\phi_i(\vecrs)=(\varphi_{1,i}(\vecrs),\varphi_{2,i}(\vecrs))^T$ of the quasiparticles.
Here the Hatree-Fock potential $\Gamma(\vecr)$ and the pair potential  or the position-dependent pair gap
$\Delta(\vecr)=\del \Ecal_{pair} /\del \rhot^*(\vecr)$  are also functional of the one-body densities including
$\rhot(\vecr)$ and $\rhot^*(\vecr)$.

The one-body
density and the pair density are evaluated as  a sum of the quasiparticle wave functions:
\begin{align}
\rho(\vecr)&=\sum_i \sum_\sigma |\varphi_{2,i}(\vecrs) |^2, \\
\rhot(\vecr)&=-\frac{1}{2}\sum_i \sum_\sigma \varphi_{2,i}^*(\vecrs)\varphi_{1,i}(\vecrs).
\end{align}
 The HFB ground state $\ket{\Psi_0}$ is obtained by solving the HFB equation with an iterative procedure.
 
\noindent
\underline{U(1) gauge symmetry and its violation}

\begin{figure}[h]
\centering
\begin{minipage}{\columnwidth}
\includegraphics[width=0.52\columnwidth]{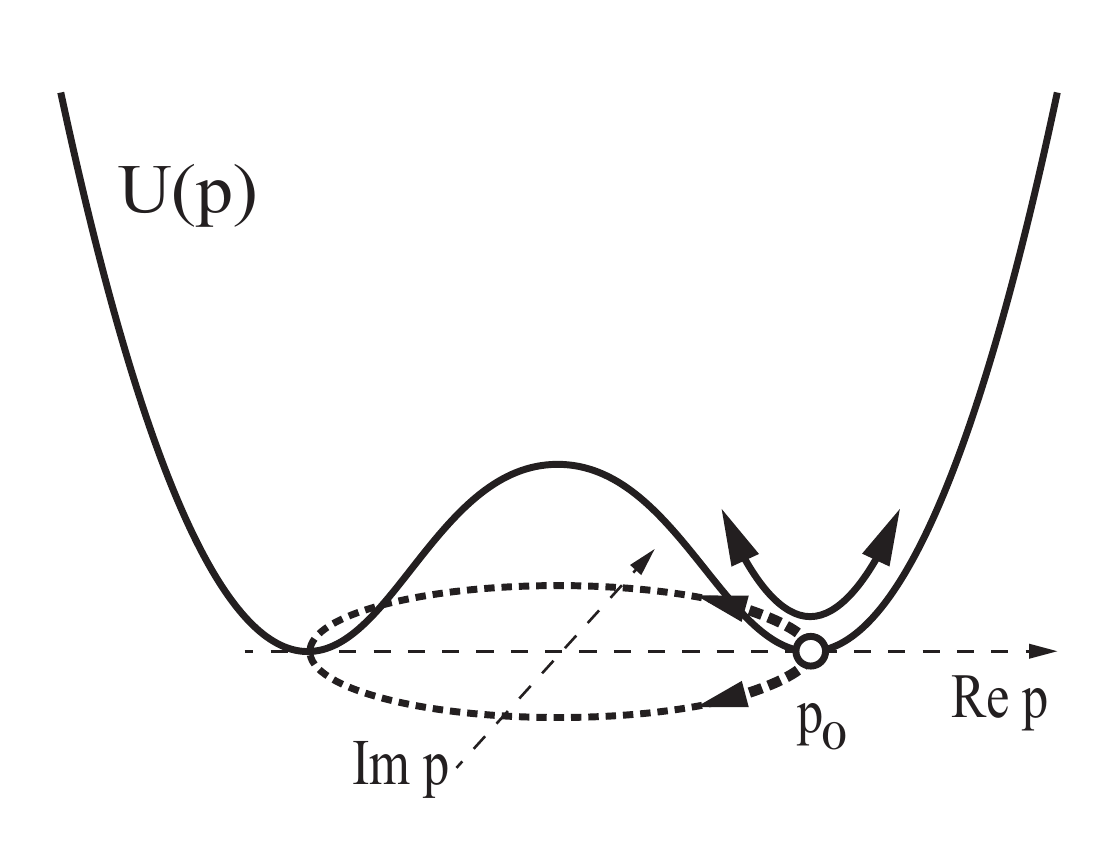}
\end{minipage} 
\caption{Schematic figure of the effective potential $U(p)=\Ecal(p)-\Ecal(0)$
as a function of the complex order parameter $p$, which 
represents the pair condensate or the pair gap. 
}
 \label{fig_mexican}
\end{figure}

For the sake of the discussions below, we briefly recapitulate the U(1) gauge symmetry in 
the Hartree-Fock-Bogoliubov (HFB) model. Consider the global U(1) gauge transformation  
$\ket{\Psi} \rightarrow e^{i\theta \hat{N}}\ket{\Psi} $ with $\hat{N}$ being the nucleon
number operator $\hat{N}=\int d\vecr \sum_{\sigma} \psi^\dagger(\vecrs)\psi(\vecrs)$.
The energy functional or the total Hamiltonian is symmetric 
with respect to the 
U(1) gauge transformation. However,
the HFB ground state $\ket{\Psi_0}$  violates
spontaneously this symmetry as
$e^{i\theta \hat{N}}\ket{\Psi_0} $ differs from $\ket{\Psi_0}$. Correspondingly
the pair condensate and the pair field 
are modulated in their phases as
\begin{align} \label{gauge-trans}
\rhot(\vecr)\rightarrow e^{2i\theta}\rhot(\vecr) , \ \ \ \ \
\Delta(\vecr) \rightarrow e^{2i\theta} \Delta(\vecr).
\end{align}
All the states $e^{i\theta \hat{N}}\ket{\Psi_0} $ transformed from one realization of the HFB ground state 
$\ket{\Psi_0} $  
are also degenerate HFB ground state.  This can be seen in the fact that the above equations are invariant
for 
$\varphi_{1,i}(\vecrs) \rightarrow e^{i\theta}\varphi_{1,i}(\vecrs)$, $\varphi_{2,i}(\vecrs) \rightarrow e^{-i\theta}\varphi_{2,i}(\vecrs)$ together with Eq.(\ref{gauge-trans}).
In the following we choose one of the HFB ground states in which  
the pair condensate $\rhot(\vecr)$ is real. In fact, the pair density and the pair potential
are assumed to be real in our numerical code for the HFB ground state. 

The situation is illustrated schematically in Fig.\ref{fig_mexican}. Here the energy gain or the
effective potential $U(p)=\Ecal(p)-\Ecal(0)$ is shown
as a function of an "order parameter" $p$, which could be an average
value of the pair condensate $\rhot(\vecr)$ or the pair potential $\Delta(\vecr)$
(Details  will be specified later). The order parameter $p$ is a complex variable
and it transforms as
$p \rightarrow e^{2i\theta}p$ under the U(1) gauge transformation. The origin
$p=0$ corresponds to the Hartree-Fock ground state where the pair condensate is
absent, and $p=p_0$ corresponds to the HFB ground state. 
The difference $U_{\mathrm{cond}}=\Ecal(p_0)-\Ecal(0)$ is the energy gain associated with the 
condensation of the Cooper pairs.

\noindent
\underline{Excitation modes}
 
 We describe excitation modes build on top of the HFB ground state $\ket{\Psi_0}$ by means of the
 quasi-particle random-phase approximation (QRPA).  The QRPA is equivalent to describing
 a linear response under an external perturbation in the frame work of the 
 time-dependent HFB theory\cite{Matsuo2001,Nakatsukasa-etal2016}.

 We consider a time-dependent perturbation $\mu e^{-i\omega t}\hat{V}_{\mathrm{ext}}$ where a perturbing field 
 $\hat{V}_{\mathrm{ext}}$ 
 belongs to a class of generalized one-body operators 
  expressed in terms of 
 the local density $\hat{\rho}(\vecr)$
 and the pair densities $\hat{P}(\vecr)$ and $\hat{P}^\dagger(\vecr)$:
 \beq \label{ext-field}
\hat{V}_{\mathrm{ext}}=\int  d\vecr \left\{ 
 v_{0}(\vecr)\hat{\rho}(\vecr) +v_{r}(\vecr)\hat{P}(\vecr) +v_{a}(\vecr)\hat{P}^\dagger(\vecr) 
 \right\}.
 \eeq
 The perturbation causes time-dependent fluctuations 
 $\delta\rho(\vecr,\omega)$, $\delta\rhot(\vecr,\omega)$ and $\delta\rhot^\star(\vecr,\omega)$
(expressed in the frequency domain) in the three densities $\rho(\vecr)$, $\rhot(\vecr)$ and $\rhot^*(\vecr)$.
 It induces also fluctuations in the self-consistent field
 \beq \label{sc-field}
 \delta \hat{V}_{sc}(\omega) = \int  d\vecr \left\{ 
 \delta\Gamma(\vecr,\omega)\hat{\rho}(\vecr) +\delta\Delta^*(\vecr,\omega)\hat{P}(\vecr) +\delta\Delta(\vecr,\omega)\hat{P}^\dagger(\vecr) 
 \right\}
 \eeq
 through the densities.  
Here we  assume that  fluctuations in the Hartree-Fock potential
 and the pair potential, $\delta\Gamma$ and $\delta\Delta$,
  are proportional to the density 
fluctuations $\delta\rho(\vecr,\omega)$, $\delta\rhot(\vecr,\omega)$ and $\delta\rhot^\star(\vecr,\omega)$.
Then 
 the linear response of the system is governed by 
  \begin{equation} \label{rpa}
\left(
\begin{array}{c}
\delta\rho(\vecr,\omega) \\
\delta\rhot(\vecr,\omega) \\
\delta\rhot^\star(\vecr,\omega) 
\end{array}
\right)
=\int_0  d\vecr'
\left(
\begin{array}{ccc}
& & \\
& R_{0}^{\alpha\beta}(\vecr,\vecr',\omega)& \\
& & 
\end{array}
\right)
\left(
\begin{array}{l}
\frac{\delta \Gamma}{\delta\rho} 
\delta\rho(\vecr',\omega)+ v_{0}(\vecr') \\
\frac{\delta\Delta^*}{\delta\rhot^*} \delta\rhot^\star(\vecr',\omega) + v_{r}(\vecr') \\
\frac{\delta\Delta}{\delta\rhot} \delta\rhot(\vecr',\omega)+ v_{a}(\vecr')
\end{array}
\right).
\end{equation}
 Here $R_{0}^{\alpha\beta}(\vecr,\vecr',\omega)$ is unperturbed density response function
 \beq \label{resp-fn}
 R_{0}^{\alpha\beta}(\vecr,\vecr',\omega)=\frac{1}{2}\sum_{ij}\left\{
 \frac{\braket{0}{\hat{\rho}_\alpha(\vecr)}{ij}\braket{ij}{\hat{\rho}_\beta(\vecr')}{0}}
 {\hbar\omega+i\eps-E_i-E_j}
 -
 \frac{\braket{0}{\hat{\rho}_\beta(\vecr')}{ij}\braket{ij}{\hat{\rho}_\alpha(\vecr)}{0}}
 {\hbar\omega+i\eps+E_i+E_j}
 \right\}
 \eeq
 with $E_i$ in the denominator being the excitation energy of the quasiparticle state $i$.
 The numerators in the r.h.s. are matrix elements
 of the density operators 
   $\hat{\rho}_\alpha(\vecr)=\hat{\rho}(\vecr),\hat{P}(\vecr),\hat{P}^\dagger(\vecr)$
   between the HFB ground state $\ket{0}=\ket{\Psi_0}$ and 
 two-quasiparticle configurations $\ket{ij}=\beta_i^\dagger\beta_j^\dagger\ket{\Psi_0}$.
Here we put  the spectral representation of the unperturbed
density response function expressed with discretized spectrum even for unbound quasiparticle
states. The continuum nature of the unbound quasiparticle states can be taken into account
explicitly using the method of the continuum QRPA formalism\cite{Matsuo2001}. 
See Appendix for the two-quasiparticle matrix elements and the continuum QRPA.

 The excited states $\ket{\nu}$ of the system appear as poles of the density fluctuations as a  
 function of the excitation energy $\hbar\omega$. The transition matrix elements for the perturbing field is
 obtained through the strength function
\beq
 S(\hbar\omega)\equiv \sum_\nu \left| \braket{\nu}{\hat{V}_{\mathrm{ext}}}{0} \right|^2 \delta(\hbar\omega - E_\nu), 
 \eeq
 which can be calculated in terms of the solution of the linear response equation:
 \begin{align}
 S(\hbar\omega) & = -\frac{1}{\pi} \mathrm{Im} \langle \hat{V}_{\mathrm{ext}}^\dagger \rangle(\omega) \nonumber \\
  & = -\frac{1}{\pi} \mathrm{Im} 
 \int d\vecr \left\{  v_{0}^*(\vecr)\delta\rho(\vecr,\omega)
 +v_{a}^*(\vecr)\delta\tilde{\rho}(\vecr,\omega)+v_{r}^*(\vecr)\delta\tilde{\rho}^{\star}(\vecr,\omega) \right\}.
 \end{align}

\noindent
\underline{Model parameters and numerical details}

We discuss the pairing correlation of neutrons in semi-magic Sn isotopes, for which
the ground states are expected to be spherical. We
adopt the SLy4 parameter set\cite{chabanat98} for the Skyrme energy functional. The pairing interaction defining the pairing energy functional is the density dependent
delta interaction (DDDI), the contact force whose interaction strength depends on the position through the
nucleon density:
\begin{align}\label{DDDI}
v^{{\rm pair}}_q(\vecr,\vecr')&={1\over2}(1-P_\sigma)V_q(\vecr)
\delta(\vecr-\vecr'), \ \ \ (q=n,p) \\
V_q(\vecr)&=v_0\left(1-\eta\left({\rho_q(\vecr)\over \rho_0}\right)^\alpha\right).
\end{align}
Correspondingly the pair potential is given by
$
\Delta_q(\vecr)=V_q(\vecr)\rhot_q(\vecr).
$
 In the actual calculation, we assume the spherical HFB mean-field and solve all the equations using
 the polar coordinate system and the partial wave expansion. The radial coordinate is truncated at
 $r=R_{\mathrm{max}}$ with $R_{\mathrm{max}}=20$fm. The partial waves are taken into account up to the maximum angular quantum
number $l_{\mathrm{max}},j_{\mathrm{max}}=12,25/2$. We truncate the quasiparticle states by maximum quasiparticle energy $E_{\mathrm{max}}=60$ MeV.  The radial coordinate is discretized with an interval $\Delta r=0.2$fm. 
The parameters of the DDDI are $v_0=-458.4$ MeV fm$^{-3}$, $\rho_0=0.08$ fm$^{-3}$, $\alpha=0.845$, and $\eta=0.71$ which are
chosen to reproduce the scattering length of the nuclear force in the $^{1}S_0$ channel and the experimental gap in $^{120}$Sn\cite{Matsuo2010}.  
The HFB equation is solved with the Runge-Kutta method with box boundary condition, i.e $\phi(r)=0$ at $r=R$. With this boundary condition, unbound quasiparticle states have discrete spectrum, and the resultant strength function does also. It is possible to impose the proper boundary
condition appropriate for unbound and continuum quasiparticle states using the framework of the continuum
QRPA\cite{Matsuo2001,Serizawa2009,Matsuo2010,Shimoyama2011,Shimoyama2013} (See also Appedix). 
The QRPA calculation is performed using the discretized spectral representation while the
continuum QRPA is also used in some examples. The smoothing parameter is $\eps=100$ keV.

\section{Pair vibrations and Higgs response}

\subsection{Response to pair transfer operators}

\begin{figure}
\centering
\begin{minipage}{\columnwidth}
\includegraphics[width=0.82\columnwidth]{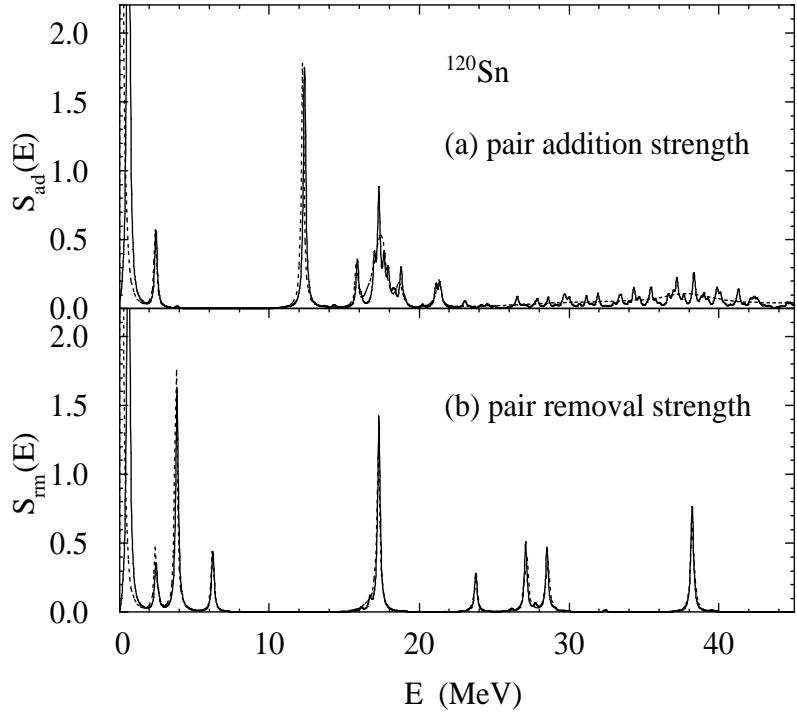}
\end{minipage} 
\caption{Pair-addition and pair-removal strength functions,
$S_{\mathrm{ad}}(E)$ and $S_{\mathrm{rm}}(E)$, for neutrons in $^{120}$Sn, plotted
as functions of the excitation  energy $E$. The unit of the strength functions is
MeV$^{-1}$ since $\Phat_{\mathrm{ad}}$ and $\Phat_{\mathrm{rm}}$ are dimensionless. 
The solid line is the result
of the QRPA calculation based on the discretized spectral representation
of the response function while the dotted line is the result of the 
continuum QRPA. The smoothing width is $\epsilon=100$ keV.}
 \label{Pad-rm-strength}
\end{figure}

 \begin{figure}
\centering
\begin{minipage}{\columnwidth}
\includegraphics[width=0.52\columnwidth]{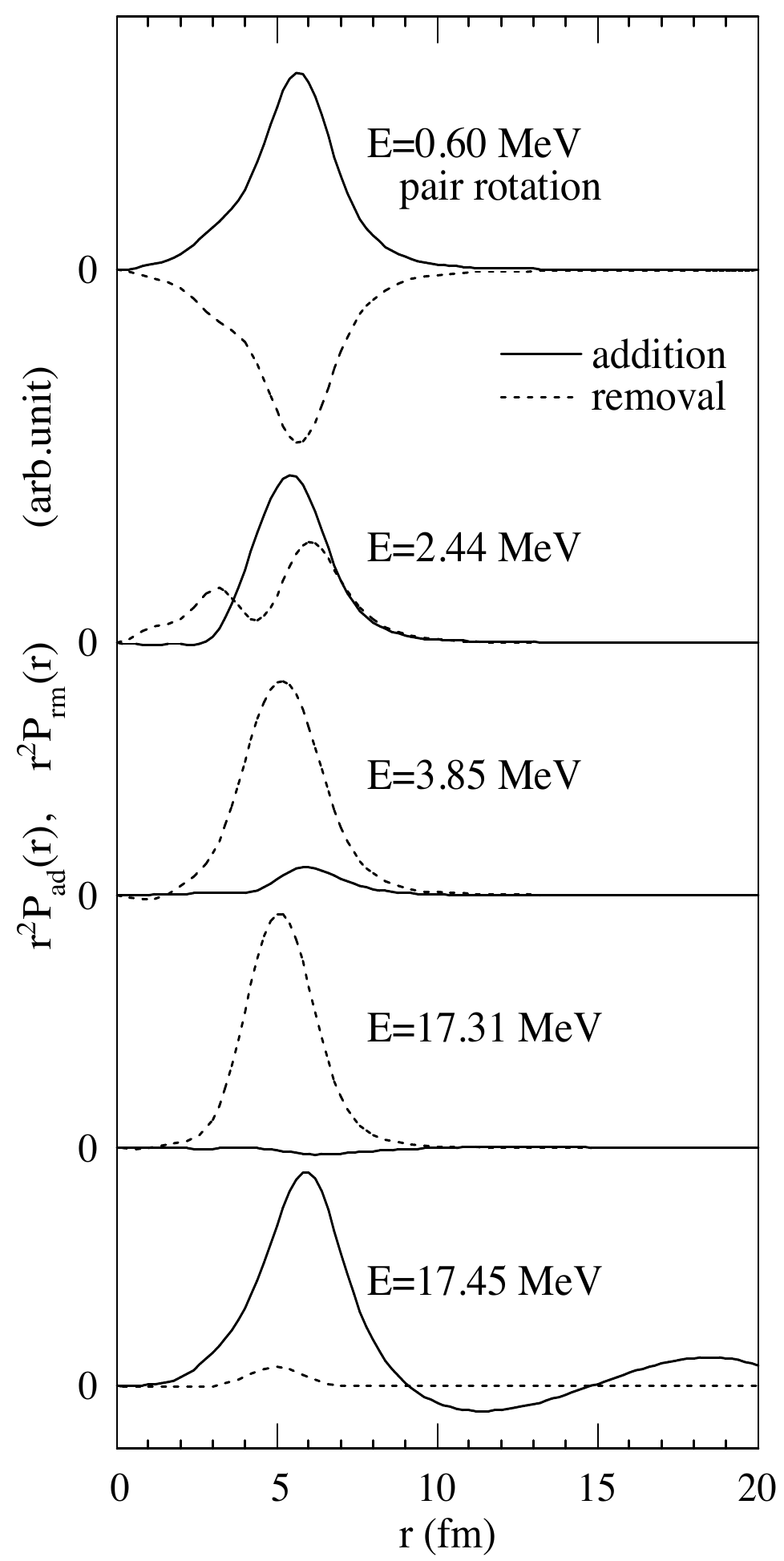}
\end{minipage} 
\caption{Pair-addition and pair-removal transition densities
$r^2P^{(\mathrm{ad})}_{\nu}(r)$ (solid curve) and $r^2P^{(\mathrm{rm})}_{\nu}(r)$ (dotted curve) of neutrons 
for some peaks seen in the pair-addition and 
pair-removal strength functions for $^{120}$Sn, shown in Fig.\ref{Pad-rm-strength}.
}
 \label{trans-dens}
\end{figure}

Sensitive probes to the  pairing correlation may be expressed in terms of the
pair field operators   $\hat{P}(\vecr) =\psi(\vecr\up) \psi(\vecr\down)$ and 
$\hat{P}^\dagger(\vecr) =\psi^\dagger(\vecr\down) \psi^\dagger(\vecr\up)$. We define
a pair-addition operator
\beq \label{Pad-op}
\hat{P}_{\mathrm{ad}}=\int d\vecr Y_{00}f(r)\hat{P}^{\dagger}(\vecr)
=\frac{1}{\sqrt{4\pi}}\int d\vecr f(r)  \psi^\dagger(\vecr \down)\psi^\dagger(\vecr \up)
\eeq
and a pair-removal operator
\beq \label{Prm-op}
\hat{P}_{\mathrm{rm}}=\int d\vecr Y_{00} f(r) \hat{P}(\vecr)
=\frac{1}{\sqrt{4\pi}}\int d\vecr f(r)  \psi(\vecr \up)\psi(\vecr \down).
\eeq
These operators bring about a transition changing particle number
by $\Delta N=\pm 2$.  
  Here we introduce a form factor $f(r)$ which is effective  in a spatial region where the
 nucleon density is finite, but diminishes far outside the nucleus as we are interested in a process where a nucleon
 pair  is added to or removed from a nucleus.  Specifically
 we choose a Woods-Saxon function, 
 \beq
 f(r)=\frac{1}{1+\exp((r-R)/a)}
\eeq
with  $R=1.27 \times A^{1/3}$ fm and $a=0.67$ fm,
 but as shown later the main conclusion does not depend on
 detailed form of $f(r)$. 
 In the present study we describe the pair vibration with the lowest multipolarity
 $J^\pi=0^+$. 
 Here $Y_{00}$ is the spherical harmonics with rank 0
 and both $\Phat_{\mathrm{ad}}$ and $\Phat_{\mathrm{rm}}$ carry the angular and parity
 quantum numbers $0^+$.
 We describe the pair vibration with spin parity
 $J^\pi=0^+$. 
 
 Response of a nucleus against these operators are represented by the strength functions 
 \beq
 S_{\mathrm{ad}}(E)=\sum_\nu \left| \braket{\nu}{\hat{P}_{\mathrm{ad}}}{0} \right|^2 \delta(E - E_\nu)
 \eeq
 for the pair-addition process, and
 \beq
 S_{\mathrm{rm}}(E)=\sum_\nu \left| \braket{\nu}{\hat{P}_{\mathrm{rm}}}{0} \right|^2 \delta(E- E_\nu)
 \eeq
for the pair-removal process. Here  $\ket{\nu}$ is the QRPA excited states
 with excitation energy $E_\nu$
whereas $\braket{\nu}{\hat{P}_{\mathrm{ad}}}{0}$ and $ \braket{\nu}{\hat{P}_{\mathrm{rm}}}{0}$ is the matrix  elements of 
the pair-add and pair-removal operators between the HFB ground state $\ket{0}=\ket{\Psi_0}$ and the 
QRPA excited states.  Transition densities
\begin{align}
P^{(\mathrm{ad})}_{\nu}(\vecr)&=\braket{\nu}{\Phat^\dagger(\vecr)}{0}=\braket{\nu}{\psi^\dagger(\vecr\downarrow)\psi^\dagger(\vecr\uparrow)}{0} 
=P^{(\mathrm{ad})}_{\nu}(r)Y_{00}^* , \\
P^{(\mathrm{rm})}_{\nu}(\vecr)&=\braket{\nu}{\Phat(\vecr)}{0}=\braket{\nu}{\psi(\vecr\uparrow)\psi(\vecr\downarrow)}{0}=P^{(\mathrm{rm})}_{\nu}(r)Y_{00}^* ,
\end{align}
representing amplitudes of pair-addition and -removal are also useful measures to look into 
structure of the excited states.

 Figure \ref{Pad-rm-strength} (a) and (b) show the  pair-add and pair-removal strength functions, respectively, 
 calculated for neutrons in $^{120}$Sn.
 The results exhibit several noticeable peaks. We first discuss them referring to  
the concept of the pair rotation and the pair vibration.
 
  First, we note the 
 peak with the lowest excitation energy  $E=0.60$ MeV, which is seen both in the pair-addition and in the pair-removal strength functions. 
 It corresponds to the pair rotation, i.e. the transition from the ground state of the system $N$ to  the ground state with
 $N \pm 2$.  The pair rotation, which is associated with a displacement along the dashed line in Fig.\ref{fig_mexican}, should appear as a zero-energy Nambu-Goldstone mode arising from the
 the gauge transformation $e^{i\theta \hat{N}}\ket{\Psi_0}$.   In the actual numerical implementation 
 of the HFB+QRPA formalism,
the calculated  excitation energy of the pair rotation deviates from zero by a small amount,  0.60 MeV
in the present case, due to numerical errors and truncations.
 The transition densities of the
lowest energy peak, the pair rotation mode, are shown in Fig.\ref{trans-dens}. It is seen that the pair-addition
transition density and the pair-removal one has the same shape as functions of $r$, but they
have the opposite phases. 
This is what is expected for the pair rotation, since 
the phase change in the pair condensate $\tilde{\rho}(\vecr)$ gives 
 $\delta\rhot(\vecr)= \delta e^{2i\delta\theta}\rhot(\vecr)\sim 2i\delta\theta\rhot(\vecr)$ and $\delta\rhot^*(\vecr)=
 \delta e^{-2i\delta\theta}\rhot(\vecr)\sim -2i\delta\theta\rhot(\vecr)$, i.e., 
 opposite sign in these quantities 
with a common shape proportional to the ground state pair condensate $\rhot(\vecr)$. 

Other peaks in Fig.\ref{Pad-rm-strength}(a) and (b) represent transitions to excited $0^+$ states in the neighbor nuclei.
The peak at $E=2.44$ MeV, the lowest excited $0_2^+$ state, has transition
strengths which are not very large. 
This state corresponds to the pair vibration, 
which is predicted to be
a lowest excited  $0_2^+$ state with energy  about $2\Delta$\cite{Bes-Broglia66,Broglia73,Brink-Broglia}.
 The excitation energy 2.44 MeV is consistent with the
predicted value $E \approx 2\Delta$ as an average value of the pair gap  is
$\Delta=\int d\vecr \rho(r)\Delta(r) /\int d\vecr\rho(r)=1.15$ MeV in the present case.

In the interval of excitation energy from 3 MeV  up to around 20 MeV, 
there exist several peaks which have larger pair-addition or -removal strengths 
than those of the pair vibration at $E=2.44$ MeV, e.g. 
 peaks at $E=12. 36$ and 17.45 MeV seen in the pair-addition strength,
 and those at $E=3.85$ and 17.31 MeV in the pair-removal
strength. These states may also be regarded as pair vibrational states as the pair-transfer
strengths are enhanced by the RPA correlation (see below). Although 
high-lying pair vibration is often called the 
giant pair vibration (GPV)\cite{Broglia1977,Cappuzzello2015,Cavallaro2019,Assie2019}, 
we adopt in the present paper
 a more inclusive  term "high-lying pair vibrations" for these peaks
since the strength distribution
does not form a single giant peak, but rather multiple peaks in a wide energy
interval.  

We remark that the low-lying pair vibration at $E=2.44$ MeV and other high-lying pair vibrations have
slightly different characters. The low-lying pair vibration has both the pair-addition and -removal strengths
whereas the high-lying pair vibrations have either the pair-addition
strength or the pair-removal strengths. This difference is also clearly seen 
in the transition density, shown in Fig.\ref{trans-dens}: the low-lying pair vibration
at $E=2.44$ MeV has both the pair-addition and pair-removal
amplitudes. Contrastingly 
 the high-lying pair vibrations 
are seen either in the pair-addition
strength function or in the pair-removal strength function, but not in the both. This is reflected also
in the transition density. As shown in Fig.\ref{trans-dens}, 
the transition density
of the peaks at $E =12.31$ MeV and
$E \approx 17.45$ MeV seen in the pair-addition
strength function have large pair-addition amplitude while the pair-removal
amplitude is almost negligible.  These states can be populated only by the
pair-addition process. On the contrary, the peaks at $E=3.85$ and 17.31 MeV existing
in the pair-removal strength function have dominant amplitude
for the pair-removal transition density.

\subsection{Higgs response}

\begin{figure}
\centering
\begin{minipage}{\columnwidth}
\includegraphics[width=0.82\columnwidth]{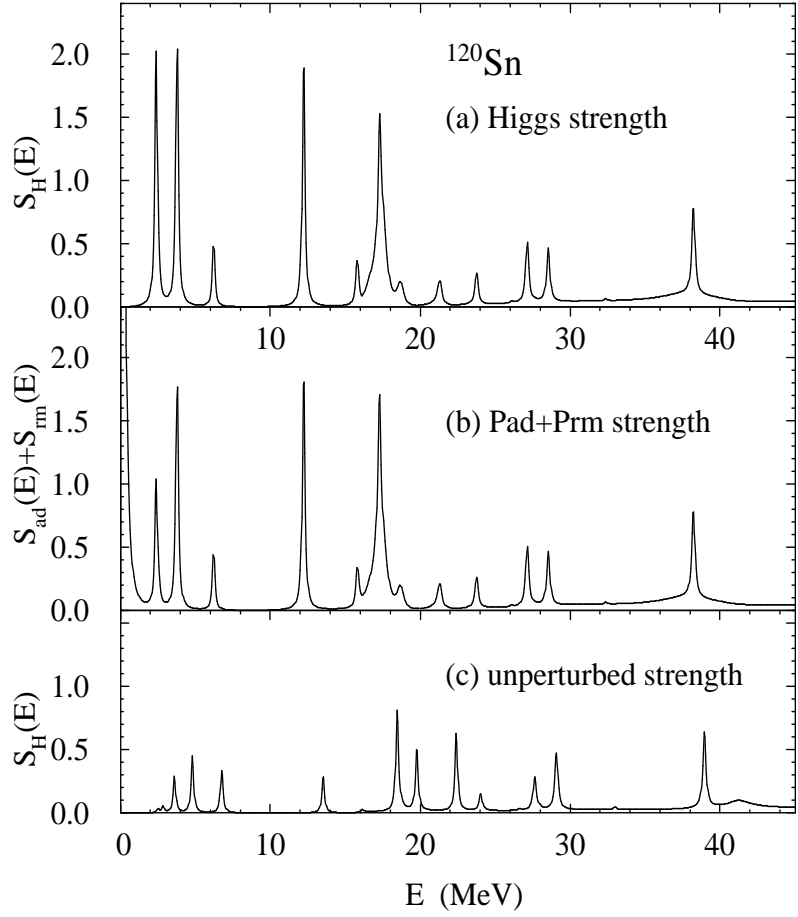}
\end{minipage} 
\caption{(a) Higgs strength function
$S_{\mathrm{H}}(E)$  of neutrons for $^{120}$Sn, plotted
as functions of the excitation  energy $E$.
(b) Sum of the 
pair-addition and pair-removal strength functions,
$S_{\mathrm{ad}}(E)+S_{\mathrm{rm}}(E)$.
(c) Unperturbed Higgs strength function $S_{\mathrm{H}}(E)$
for uncorrelated neutron two-quasiparticle excitations.
}
 \label{Higgs-strength}
\end{figure}

We shall here address a question how the pair vibrations discussed above are related to 
the macroscopic picture as an amplitude oscillation of the order parameter,
 the Higgs mode. However, the existence of multiple states, including
both the low-lying and high-lying pair vibrations,  makes it  difficult to relate 
this macroscopic picture to one of  the pair vibrational
states. We need a new way of characterizing the pair vibrations, which is
more suitable for obtaining the macroscopic information.

As is illustrated by Fig.\ref{fig_mexican}, response of the pair fields around a HFB equilibrium point has
two different directions with respect to the order parameter of the pair condensation:
the phase mode changing the
phase of the pair gap (or the pair condensate), and the other changing the
amplitude of the pair gap/condensate.  
The pair-add operators, $\Phat_{\mathrm{ad}} \sim \psi^\dagger\psi^\dagger$, and the   
pair-removal operators $\Phat_{\mathrm{rm}} \sim \psi\psi$, do not probe separately   these two
different modes since both the pair rotation (the phase
mode) and the pair vibrations (possible amplitude mode) are seen in
both of the pair-addition
and pair-removal strengths functions. 

Instead  we introduce a pair field operator that is
a linear combination of $\Phat_{\mathrm{ad}} $ and $\Phat_{\mathrm{rm}}$:
\beq \label{Higgs-op}
\hat{P}_{\mathrm{H}}= \hat{P}_{\mathrm{ad}} + \hat{P}_{\mathrm{rm}}=\frac{1}{\sqrt{4\pi}}\int d\vecr f(r) 
\left\{ \psi(\vecr \up)\psi(\vecr \down) +\psi^\dagger(\vecr \down)\psi^\dagger(\vecr \up) \right\}.
\eeq
Note that fluctuation of the amplitude $|\rhot(\vecr)|$ of the pair condensate reads
$\delta|\rhot(\vecr)|=\left(\rhot(\vecr)\delta\rhot^*(\vecr,t)+\rhot^*(\vecr)\delta\rhot(\vecr,t)\right)/2|\rhot(\vecr)|=
(\delta\rhot^*(\vecr,t)+\delta\rhot(\vecr,t))/2$ for real $\rhot(\vecr)$, and therefore the Higgs operator $\hat{P}_{\mathrm{H}}$ 
defined by Eq.(\ref{Higgs-op}) probes the amplitude fluctuation.
In the following, we call $\Phat_{\mathrm{H}}$ the Higgs operator after the nomenclature that the amplitude mode of the
pair field is often called Higgs mode\cite{Pekker-Varma,Shimano-Tsuji}. The strength function for the Higgs operator is 
defined by
\beq
S_{\mathrm{H}}(E) = \sum_\nu \left| \braket{\nu}{\hat{P}_{\mathrm{H}}}{0} \right|^2 \delta(E - E_\nu).
\eeq

Figure \ref{Higgs-strength} (a) is the calculated Higgs strength function $S_{\mathrm{H}}(E)$
of neutrons for $^{120}$Sn.
For comparison we show  in Fig.\ref{Higgs-strength}(b)  a sum 
 $S_{\mathrm{ad}}(E)+S_{\mathrm{rm}}(E)$ of the pair-addition and pair-removal
strength functions (Fig.\ref{Pad-rm-strength}). The panel (c) is a
Higgs strength function for independent two-quasiparticle excitations,
namely a calculation in which the RPA correlation is removed.
We observe the following features. 

We first note  that, compared with the uncorrelated
result (c), the Higgs strength in (a) is enhanced by several times for the peaks of 
low-lying and high-lying pair vibrations. This clearly indicates collectivity of both the
low- and high-lying pair vibrations. 

Secondly we observe that 
the Higgs strength function is close to the sum of the pair-addition and pair-removal 
strengths although there are some
differences. A significant and important difference is seen for the pair rotation mode 
(the peak at $E=0.60$ MeV in this example), for which the Higgs strength almost invisible.
Indeed, it should vanish in principle due to the very nature of the phase mode 
associated with the U(1) gauge symmetry. As discussed above, 
the pair-add and pair-removal amplitudes of the phase mode 
have the same shape but with opposite sign (cf. Fig.\ref{trans-dens}),
and consequently these two amplitudes cancel each other in the matrix element
of the Higgs operator $\braket{\nu}{\Phat_{\mathrm{H}}}{0}=\braket{\nu}{\Phat_{\mathrm{ad}}}{0} + \braket{\nu}{\Phat_{\mathrm{rm}}}{0}=0$.

For the low-lying pair vibration (the peak at $E=2.44$ MeV), contrastingly, 
the Higgs strength is enhanced. Namely it is more than the sum of the pair-addition
and pair-removal strengths (It is approximately two times the sum). This is because 
the both pair-addition and pair-removal transition densities of this mode
have the same sign, as seen in Fig.\ref{trans-dens}, and they leads to a
coherent sum in  the matrix element of the Higgs operator
$\braket{\nu}{\Phat_{\mathrm{H}}}{0} = \braket{\nu}{\Phat_{\mathrm{ad}}}{0}+ \braket{\nu}{\Phat_{\mathrm{rm}}}{0}$.
Consequently the low-lying pair vibration  contributes more strongly
to the Higgs strength than to the pair-addition and pair-removal strengths. Because of
this enhancement due to the coherence, the low-lying pair vibration could be regarded
as an Higgs mode in nuclei. This interpretation might be supported also by
the energy relation $E \approx 2\Delta$ of the low-lying pair vibration, which is
also the case for
the Higgs mode in the superconductors\cite{Anderson1958,Littlewood-Varma1981-82,Pekker-Varma,Shimano-Tsuji}
with the energy relation $E = 2\Delta$. However, for the reasons below we reserve possible identification
of the low-lying pair vibration as the Higgs mode.

 The high-lying pair vibrations, emerging as several peaks distributed  up to around $E\sim 20$ MeV,
 have the Higgs strength comparable to that of the low-lying pair vibration. From the viewpoint of
 the response to the Higgs operator, the high-lying pair vibrations may also be candidates of 
 the Higgs mode. However the energy relation $E = 2 \Delta$ does not hold.
The high-lying pair vibrations exhibit no coherence between the matrix elements 
 $\braket{\nu}{\Phat_{\mathrm{ad}}}{0}$ and $\braket{\nu}{\Phat_{\mathrm{rm}}}{0}$. 
 The Higgs strengths 
 of these states are incoherent sum of the pair-addition and -removal strengths. It may not be
 reasonable to identify either the
 high-lying or the low-lying pair vibrations as a pure Higgs mode.

 
 \subsection{Static polarizability}

The Higgs strength is not concentrated to a single state, but 
distributed over many excited states, including both the low-lying and high-lying pair vibrations.
Since there seems no one single pair vibrational state interpreted as a pure Higgs mode, we need
a different viewpoint which integrates both the low- and high-lying pair vibrations.
Here we consider the static polarizability
which can be derived from the Higgs strength function. It is defined by
\beq \label{static-pol}
\alpha_{\mathrm{H}}= \frac{d \langle \Phat_{\mathrm{H}} \rangle'}{d\mu},
\eeq
where  $\langle \Phat_{\mathrm{H}} \rangle'$ is a change of the expectation value
under the static perturbation $\hat{V}_{\mathrm{ext}}=\mu \Phat_{\mathrm{H}}$. The static
polarizability is related to the strength function through the 
inversely energy weighted sum
\beq \label{e-inv-sum}
I_{-1}=2\int_{E_{\mathrm{min}}}^{E_{\mathrm{max}}} dE \frac{S_{\mathrm{H}}(E)}{E}=2\sum_\nu \frac{\left| \braket{\nu}{\hat{P}_{\mathrm{H}}}{0} \right|^2}{E_\nu} = \alpha_{\mathrm{H}}
\eeq
of the Higgs strength function $S_{\mathrm{H}}(E)$. 
In other words, the static polarizability for the Higgs operator, called Higgs polarizablility hereafter, 
can be evaluated if the 
Higgs strength function is obtained. We note that the inversely energy-weighted sum
and the static polarizability has been discussed intensively for the case of the 
E1 strength  function, i.e. the electric dipole excitations in nuclei, as a probe of
neutron skin and nuclear matter properties 
\cite{Reinhard-Nazarewicz2010, Tamii2011,Piekarewicz2012,Roca-Maza2015,Roca-Maza-Paar2018}.

\begin{figure}
\centering
\begin{minipage}{\columnwidth}
\includegraphics[width=0.82\columnwidth]{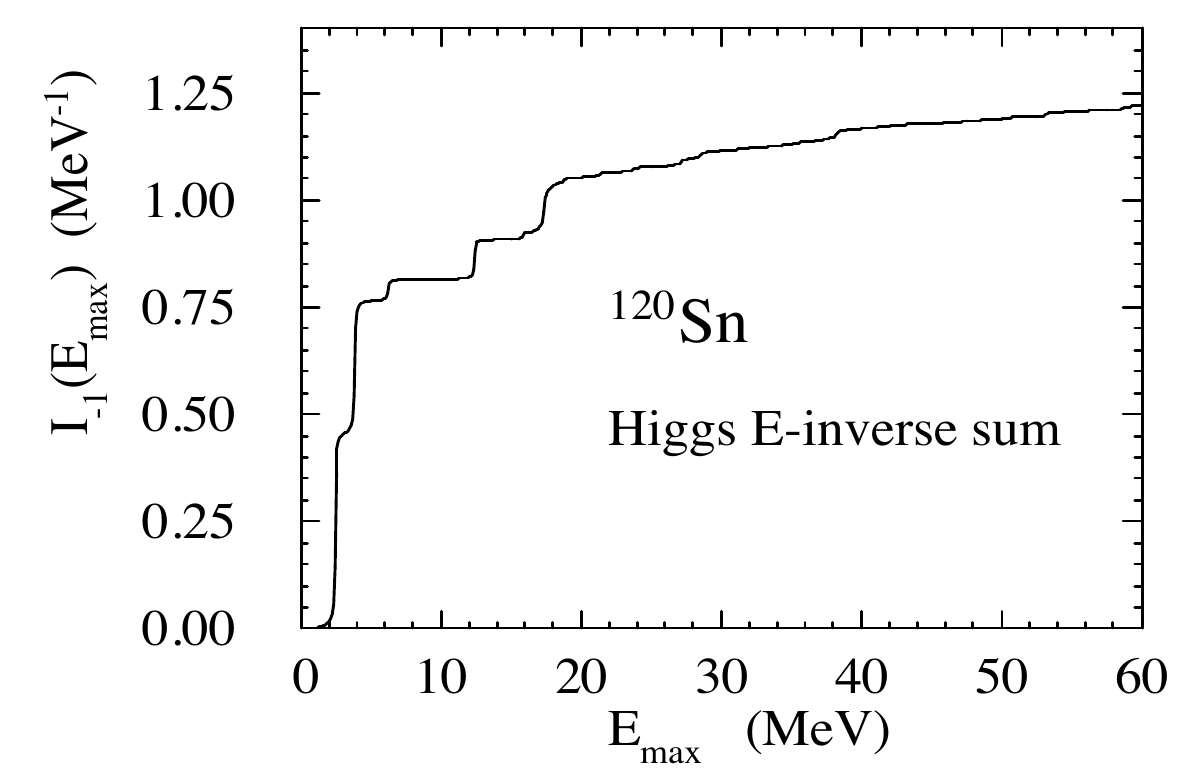}
\end{minipage} 
\caption{Inversely energy-weighted sum $I_{-1}(E_{\mathrm{max}})$ of the Higgs strength
function for $^{120}$Sn, as a function of the maximum excitation energy $E_{\mathrm{max}}$ of the
sum.   
}
 \label{fig_e_inv_sum}
\end{figure}

In the present study the inversely weighted sum is calculated by taking an integral
in an energy interval from $E_{\mathrm{min}}=1.1$ MeV to the upper limit of the excitation energy  $E_{\mathrm{max}}=60$ MeV.
The lower bound is set to exclude the contribution from the pair rotation, which should have vanished
if the complete selfconsistency is fulfilled in the numerical calculation. 
Figure \ref{fig_e_inv_sum} is a running sum of Eq.(\ref{e-inv-sum}) where the upper bound $E_{\mathrm{max}}$  is 
varied. It is seen that although the contribution of the low-lying pair vibration mode (the lowest excitation)
is as large as about 35\% of the total sum, strengths distributed at higher excitation energies contribute
significantly. The large strengths of the high-lying pairing vibrations (distributed up to around 20 MeV),
gives sizable contribution of about 45\% (i.e.  80\% including all up to 20 MeV) to the total sum. 
Contributions of strengths  above 20 MeV is small, and there is no distinct distribution
 of the strengths.

\section{Pair condensation energy}

\subsection{Effective potential}

 \begin{figure}
\centering
\begin{minipage}{\columnwidth}
\includegraphics[width=0.7\columnwidth]{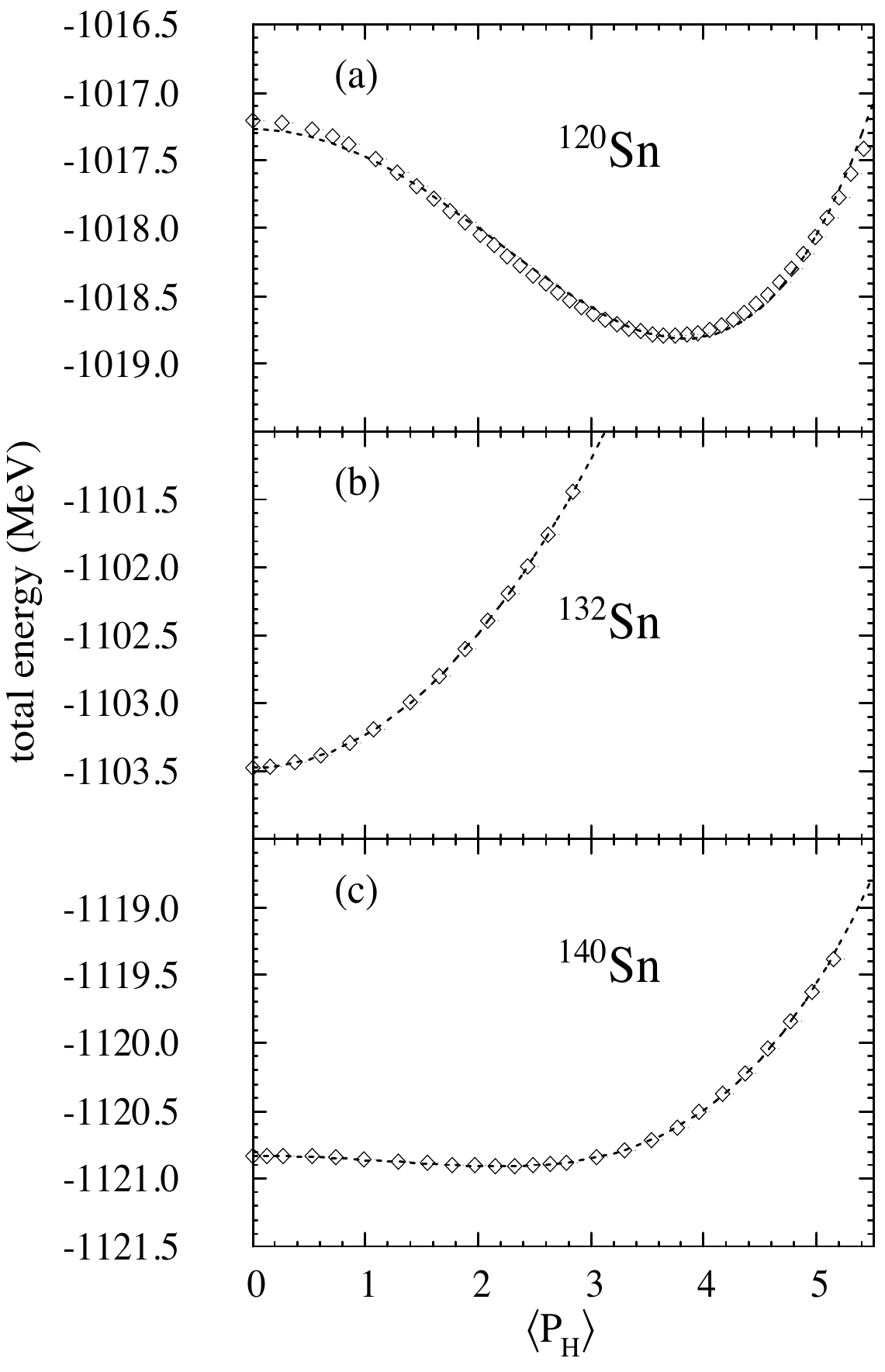}
\end{minipage} 
\caption{The potential energy curve as a function of the
order parameter $p=\langle \Phat_{\mathrm{H}} \rangle$ for $^{120}$Sn,  $^{132}$Sn
and $^{140}$Sn. The total energy obtained with the constrained Hartree-Fock-Bogoliubov (CHFB)
calculation is plotted with the diamond symbol. The dashed curve is 
a quartic function, Eq.(\ref{quartic-fn}), fitted to the CHFB results.
}
 \label{fig_cond_potential}
\end{figure}

Let us now relate the above results to the macroscopic picture discussed with Fig.\ref{fig_mexican}.
For this purpose we specify the order parameter and the effective potential.

As the order parameter we adopt the expectation value of the pair removal
operator 
\beq
p = 2\langle \Phat_{\mathrm{rm}}\rangle = \frac{1}{\sqrt{\pi}}\int d\vecr f(r) \rhot(\vecr),
\eeq
which also represents a spatial integral (or average) of the pair condensate $\rhot(\vecr)$. 
(The factor of two is put here just for later convenience). The order parameter $p$  is a 
complex variable as
it follows $p \rightarrow e^{2i\theta}p$ under the U(1) gauge transformation. We then
consider 
the total energy of the system $\Ecal(p)$ as the effective potential, a function of the order parameter $p$.
Because of the U(1) symmetry $\Ecal(p)$ depends only on the amplitude $|p|$, and it is
sufficient to consider the potential curve along the real axis of $p$, which 
corresponds to the expectation value of the Higgs operator $\Phat_{\mathrm{H}}$.
 In the following analysis we use
\beq
 p = \langle \Phat_{\mathrm{H}} \rangle
\eeq
which in fact represents  the amplitude $|p|$ of the order parameter.

The effective potential $U(p)=\Ecal(p)-\Ecal(0)$  can be calculated with the total energy $\Ecal(p)$ 
for the generalized Slater determinant $\ket{\Psi(p)}$, which is obtained
by means of the constrained Hartree-Fock-Bogoliubov (CHFB) calculation\cite{Ring-Schuck}
 using $\Phat_{\mathrm{H}}$ as a constraining operator. 
 
Figure \ref{fig_cond_potential} shows the  effective potential calculated
for $^{120}$Sn, $^{132}$Sn and $^{140}$Sn.
Let us focus on the representative example of $^{120}$Sn. 
The effective potential in this example has a so-called Mexican hat shape, which has a minimum
at finite $p=p_0$ where $p_0=\langle \Phat_{\mathrm{H}} \rangle$ corresponds 
to the HFB ground state $\ket{\Psi_0}$. Since the shape is
smooth as a function of $p$, it may be approximated by a polynomial. Guided by the
Ginzburg-Landau theory of the superconductivity\cite{Ginzburg-Landau}, we here assume a
fourth-order polynomial
\beq \label{quartic-fn}
U_{\mathrm{4th}}(p) = ap^4 +b p^2+c,
\eeq
and we fit this function to the calculated effective potential $\Ecal(p)$. As seen in Fig.\ref{fig_cond_potential}, 
the fourth-order polynomial is a good representation of the CHFB results.
We have checked and confirmed good accuracy of Eq.(\ref{quartic-fn}) for other even-even Sn isotopes, 
see two other examples in Fig.\ref{fig_cond_potential}.

\subsection{Pair condensation energy}

In the case where the effective potential is approximated well by the fourth-order polynomial
there holds a simple relation between the potential parameters, $a$ and $b$, and  
the order parameter $p_0$ and the curvature $C$ of the effective potential  at the potential
minimum:
\begin{align}
p_0&=\sqrt{\frac{-b}{2a}}, \\
C&=\left. \frac{d^2 U_{\mathrm{4th}}(p)}{d^2 p}\right|_{p=p_0}=-4b.
\end{align}
As a consequence, the potential depth $D$ can be also related to 
the two parameters characterizing the minimum:
\beq
D=U_{\mathrm{4th}}(p_0)-U_{\mathrm{4th}}(0) = -\frac{b^2}{4a}=-\frac{1}{8}Cp_0^2.
\eeq

We remark here that the parameters $p_0$ and $C$ can be derived from the
Higgs response. The order parameter $p_0$ at the minimum is
the expectation value of the Higgs operator for the ground state
 \beq
p_0=\braket{\Psi_0}{\Phat_{\mathrm{H}}}{\Psi_0}.  
\eeq
 The potential curvature $C$ at the minimum $p=p_0$
is related to the Higgs polarizability $\alpha_{\mathrm{H}}$, Eq.(\ref{static-pol}),
\beq
C=1/\alpha_{\mathrm{H}}
 \eeq
 through the Hellmann-Feynman theorem\cite{Hellmann-Feynman,Ring-Schuck}.
 Consequently the potential depth $D$, which can be interpreted as the pair condensation
 energy $U_{\mathrm{cond}}=\Ecal(p_0)-\Ecal(0)$, can be evaluated as
 \beq \label{cond-eng}
 U_{\mathrm{cond}}^{\mathrm{Higgs}}=-\frac{1}{8}\frac{p_0^2}{\alpha_{\mathrm{H}}}
 \eeq
 using the Higgs polarizability $\alpha_{\mathrm{H}}$.
 Note that the quantities in the last expression can be evaluated using
  the matrix element of the Higgs operator $\Phat_{\mathrm{H}}$ for the
  ground state $\ket{\Psi_0}$, and the strength function
  of the Higgs operator $\Phat_{\mathrm{H}}$, i.e. the information of
  the Higgs response of the system.
  
 The pair condensation energy in $^{120}$Sn obtained using the Higgs response, i.e. Eq.(\ref{cond-eng}).
 is $U_{\mathrm{cond}}^{\mathrm{Higgs}}=1.51$ MeV.
  It is compared the pair condensation energy obtained directly from the
CHFB calculation, $U_{\mathrm{cond}}^{\mathrm{CHFB}}=1.58$ MeV.  It is found that the evaluation using the 
Higgs response reproduces the actual value (
 obtained directly from the CHFB calculation) within an error less than 10 \%.
 It suggests that the pair condensation energy may be evaluated
 via the Higgs response with  good accuracy.

  \section{Sn isotopes}
  
\begin{figure}
\centering
\begin{minipage}{\columnwidth}
\includegraphics[width=0.82\columnwidth]{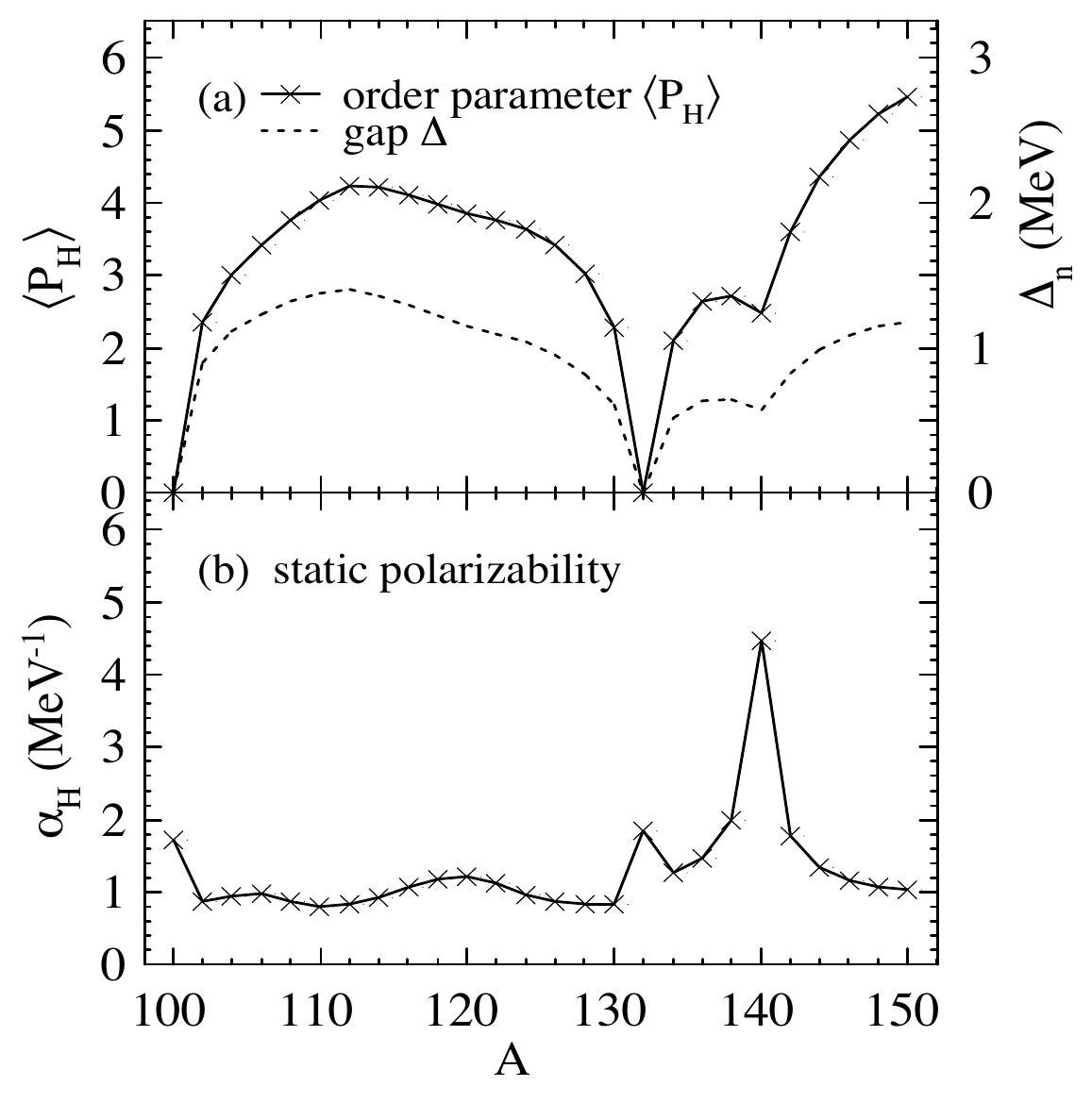}
\end{minipage} 
\caption{(a)The order parameter $p_0=\langle \Phat_{\mathrm{H}} \rangle$ and the
average pair gap $\Delta$ of neutrons, calculated for the ground states of the 
Sn isotopes.  (b) The static polarizability $\alpha_{\mathrm{H}}$ for the Higgs operator $\hat{P}_H$, calculated
from the inversely energy weighted sum of the Higgs strength function.
}
 \label{fig_order_gap_static}
\end{figure}

\begin{figure}
\centering
\begin{minipage}{\columnwidth}
\includegraphics[width=0.82\columnwidth]{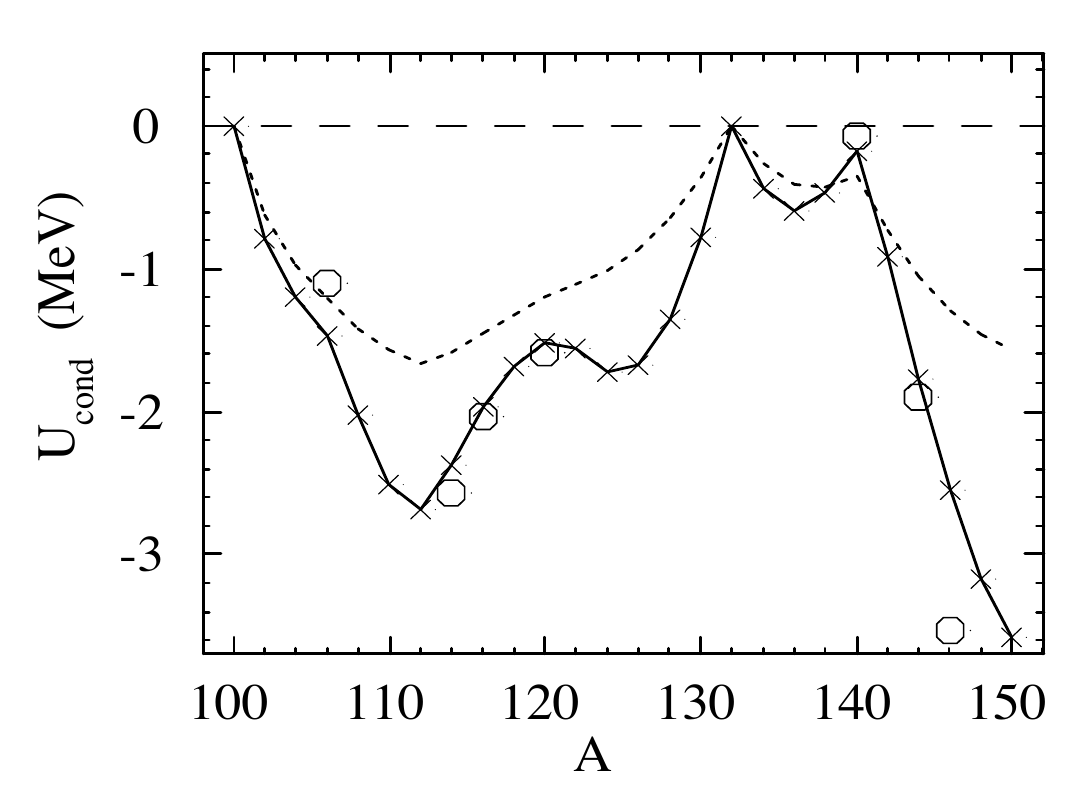}
\end{minipage} 
\caption{The pair condensation energy $U_{\mathrm{cond}}^{\mathrm{Higgs}}$ evaluated using the  
Higgs strength function in the Sn isotopes is plotted with crosses. 
It is compared with the pair condensation energy $U_{\mathrm{cond}}^{\mathrm{CHFB}}$ 
which is calculated directly using the constrained Hartree-Fock-Bogoliubov method,
plotted with circles. The dotted line is an analytic estimate
$\frac{1}{2}g\Delta^2$ with the oscillator estimate of the
single-particle level density $g=0.015A$ MeV$^{-1}$. }
 \label{fig_cond_eng}
\end{figure}

We shall apply the above method of evaluating the pair condensation energy to the neutron pair correlation
in even-even Sn isotopes covering from doubly-magic neutron-deficient $^{100}$Sn to neutron-rich 
$^{150}$Sn.
The ground state  value of the order parameter
$p_0=\langle \Phat_{\mathrm{H}} \rangle $ as well as  the average pairing gap  $\Delta=\int d\vecr \rho(r)\Delta(r) /\int d\vecr\rho(r)$
are shown in Fig.\ref{fig_order_gap_static} (a).
It is seen that these two
quantities  are 
essentially proportional to each others. They are zero in 
neutron magic nuclei $^{100}$Sn and $^{132}$Sn, where the pair condensate vanishes
due to a large shell gap at the neutron
Fermi energy, corresponding to the "normal" 
phase. 

Figure \ref{fig_order_gap_static} (b) shows the Higgs polarizability $\alpha_{\mathrm{H}}$, evaluated from the Higgs strength function.  
A noticeable feature is that its neutron number dependence is not  correlated with
that of the pairing gap $\Delta$ or the order parameter $p_0=\langle \Phat_{\mathrm{H}} \rangle$. 
For instance the Higgs 
polarizability in the closed-shell nuclei
$^{100}$Sn and $^{132}$Sn is larger than neighboring isotopes. It is also remarkable that the largest value
is in $^{140}$Sn where  the pairing gap takes an intermediate value. 

Figure \ref{fig_cond_eng} is the pair condensation energy $U_{\mathrm{cond}}^{\mathrm{Higgs}}$ of neutrons evaluated from the
Higgs response, i.e. using Eq.(\ref{cond-eng}).  It is compared with 
the  pair condensation energy $U_{\mathrm{cond}}^{\mathrm{CHFB}}$ which is calculated directly using the constrained
Hartree-Fock-Bogoliubov method for several isotopes. We confirm that
the proposed method works well not only for $^{120}$Sn discussed above,
but also for the other isotopes. 

Another point is that the isotopic trend of the pair condensation energy
shows some resemblance and difference from that of the pair gap. For comparison
we plot also an
estimate of the pair condensation energy
$\frac{1}{2}g\Delta^2$  based on the equidistant single-particle  model\cite{Ring-Schuck}
where $g$ is the single-particle level density. This estimate reflects the pair correlation through
the pair gap alone.
It is seen that the microscopic condensation energy differs from this estimate
with respect to both the magnitude and the isotopic dependence. 
This points to that the pair condensation energy is a physical quantity which carries 
 information  not necessarily expected from the pairing gap.  

\begin{figure}
\centering
\begin{minipage}{\columnwidth}
\includegraphics[width=0.82\columnwidth]{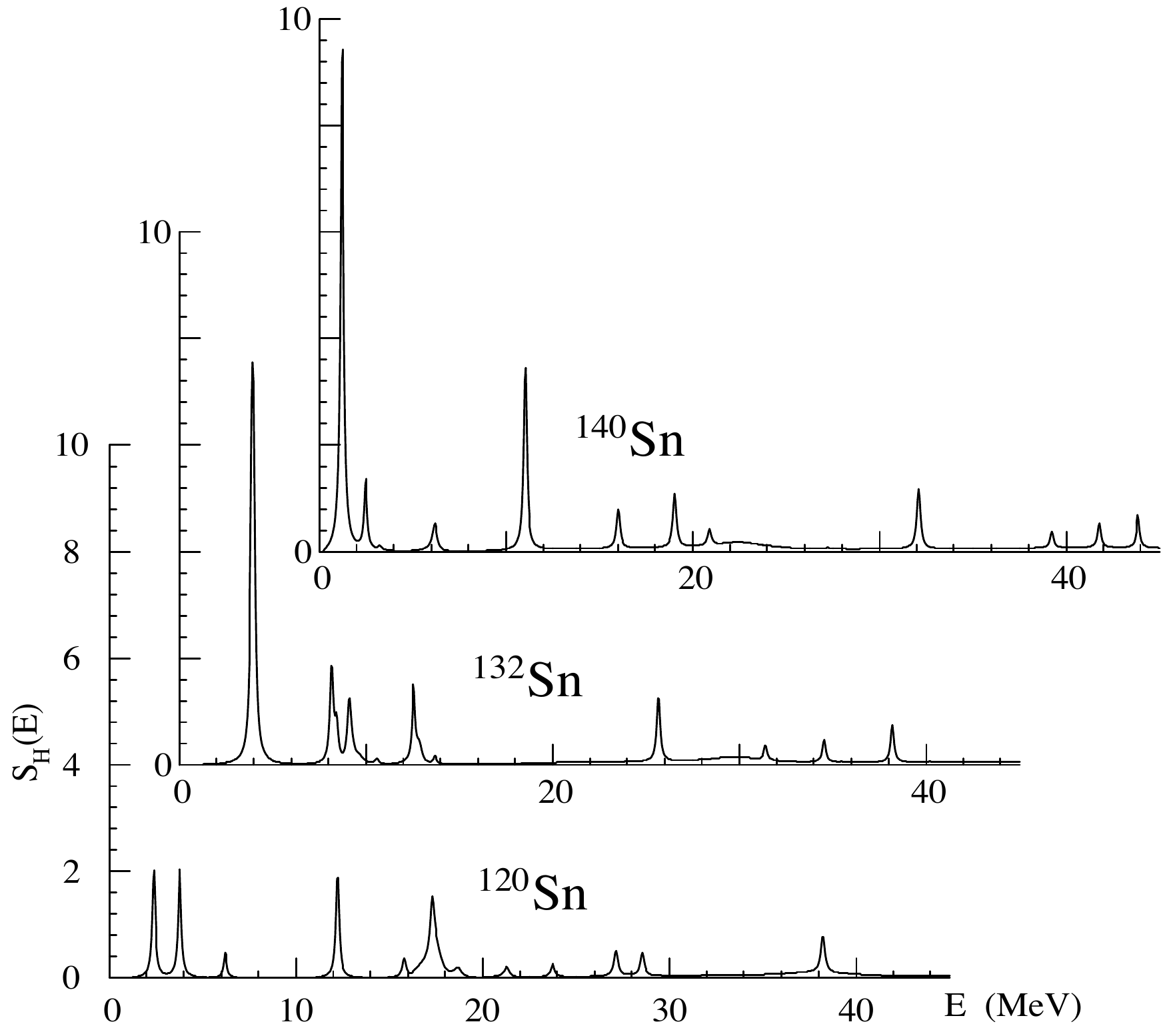}
\end{minipage} 
\caption{Higgs strength function $S_{\mathrm{H}}(E)$ for $^{120}$Sn, $^{132}$Sn and $^{140}$Sn,
calculated with the continuum QRPA. 
}
 \label{fig_higgs_120-132-140}
\end{figure}

A noticeable example is $^{140}$Sn, where the pair condensation energy is very small
(about 0.1 MeV).  The small condensation energy is due to the large value of the
Higgs polarizability (see Fig.\ref{fig_order_gap_static} (c) ), and its origin is seen 
in the Higgs strength function, shown in Fig.\ref{fig_higgs_120-132-140}.
The small pair condensation energy, reflecting the very shallow effective potential  
with small potential curvature (Fig.\ref{fig_cond_potential}(a)),
manifests itself as the presence of the low-lying pair vibration which has a very large Higgs strength,
several times larger than that in $^{120}$Sn.

We also note $^{132}$Sn, in which the HFB ground state has no pair condensate.
Correspondingly the effective potential  shown in Fig. \ref{fig_cond_potential} (b) has the minimum at $p_0=0$,
and there is no pair rotation in this case. 
The Higgs polarizability (or the potential curvature at the minimum) in this case is slightly larger 
(smaller) than that for $^{120}$Sn 
 (see Fig.\ref{fig_order_gap_static} (b) and Fig.\ref{fig_cond_potential} ). This feature can be traced to
 the presence of the pair vibrations with relatively large pair-transfer strengths as seen
 in Fig.\ref{fig_higgs_120-132-140}.

 \section{Discussion}
 
 The method to evaluate the pair condensation energy from the Higgs response
should not depend on details of the definitions of the pair transfer operators
$\Phat_{\mathrm{ad}}$, $\Phat_{\mathrm{rm}}$ and $\Phat_{\mathrm{H}}$.
We have checked other choices of the form factor $f(r)$ appearing in Eqs.(\ref{Pad-op}) and (\ref{Prm-op})
by using $f(r)=\frac{d}{dr}\frac{1}{1+\exp((r-R)/a)}$ and $f(r)=\frac{r^2}{1+\exp((r-R)/a)}$,
both of which have a weight on the surface area of the nucleus. The pair condensation energy
derived from the Higgs response in $^{120}$Sn is $U^{\mathrm{Higgs}}_{\mathrm{cond}}=1.51$ MeV and $1.58$ MeV
for the above two choices. They coincide well with $U^{\mathrm{Higgs}}_{\mathrm{cond}}=1.51$ MeV obtained with
the volume-type form factor $f(r)=\frac{1}{1+\exp((r-R)/a)}$. 
We also note that Eq. (\ref{cond-eng}) for the pair condensation energy
is expressed with a relative quantity $\alpha_{\mathrm{H}}/p_0^2$, 
for which relevant are not the absolute value but 
relative transition strengths $S_{\mathrm{H}}(E)/p_0^2$ or 
$\braket{\nu}{\Phat_{\mathrm{H}}}{0}^{2}/\langle \Phat_{\mathrm{H}} \rangle^2$ normalized
by the ground state matrix element $p_0=\langle \Phat_{\mathrm{H}} \rangle$.

We envisage application of the present method to experimental studies
using  two-neutron transfer reactions, such as $(p,t)$ and
$(t,p)$ reactions.  As discussed in Ref.\cite{Yoshida62,Broglia73,Brink-Broglia}, the cross sections of a two-neutron
stripping and pick-up reactions may be related to matrix elements
of the pair-addition
and the pair-removal operators if one assumes a one-step process.   In this simple picture, 
the transition amplitude for the ground-to-ground transition $N \rightarrow N\pm 2$
 may correspond to
matrix elements $\braket{0_{\mathrm{gs}}^+, N+2}{\Phat_{\mathrm{ad}}}{0_{\mathrm{gs}}^+, N}$
 and $\braket{0_{\mathrm{gs}}^+, N-2}{\Phat_{\mathrm{rm}}}{0_{\mathrm{gs}}^+, N}$, which is approximated by
 the expectation value $\braket{\Psi_0}{\Phat_{\mathrm{ad}}}{\Psi_0}=\braket{\Psi_0}{\Phat_{\mathrm{rm}}}{\Psi_0}=\langle \Phat_{\mathrm{H}} \rangle/2$ 
for the HFB ground state $\ket{\Psi_0}$.
 Thus the ground-to-ground pair transfer is expected to provide the order parameter $p_0=\langle \Phat_{\mathrm{H}} \rangle$.
 Similarly the two-neutron transfer cross sections populating excited $0^+$ states
 may be  related to matrix elements $\braket{0_{\nu}^+, N+2}{\Phat_{\mathrm{ad}}}{0_{\mathrm{gs}}^+, N}$
 and $\braket{0_{\nu}^+, N-2}{\Phat_{\mathrm{rm}}}{0_{\mathrm{gs}}^+, N}$, which correspond to 
 the matrix elements $\braket{\nu}{\Phat_{\mathrm{ad}}}{0}$ and $\braket{\nu}{\Phat_{\mathrm{rm}}}{0}$
 for the QRPA excited states $\ket{\nu}$. We thus expect that
 two-neutron stripping and pick-up reactions populating excited $0^+$ states
 provides the pair-addition and pair-removal strength functions, $S_{\mathrm{ad}}(E)$
 and $S_{\mathrm{rm}}(E)$. The Higgs strength function $S_{\mathrm{H}}(E)$ is then evaluated as a
 sum of $S_{\mathrm{ad}}(E)$ and $S_{\mathrm{rm}}(E)$ (cf. Fig.\ref{Higgs-strength}). Here
 transition to the low-lying pair vibration needs to be treated separately
 since for this case the Higgs transition amplitude is a coherent
 sum of the amplitudes  $\braket{\nu}{\Phat_{\mathrm{ad}}}{0}$ and $\braket{\nu}{\Phat_{\mathrm{rm}}}{0}$.
 Once the Higgs strength function is obtained in this way, the Higgs polarizability $\alpha_{\mathrm{H}}$
 can be evaluated. Consequently, combining $p_0$ and $\alpha_{\mathrm{H}}$ thus obtained, we
 may estimate the neutron pair condensation energy $U_{\mathrm{cond}}^{\mathrm{Higgs}}$ using Eq.(\ref{cond-eng}).
 Note that 
this argument based on the one-step  picture may be simplistic from a viewpoint of the
 reaction mechanisms, such as the finite range effect and the two-step processes\cite{Potel2013}. More quantitative analysis of the  two-nucleon transfer reactions needs to be explored in detail. However
it is beyond the scope of the present work, and we leave it for a future
 study. 
 
 \section{Conclusions}
 
 We have discussed a new idea  of the Higgs response, which probes the Cooper pair condensate
  in pair correlated nuclei. It is based on the analogy between the pair correlated nuclei and the 
  superfluid/superconducting state in infinite Fermi systems. The latter systems  exhibit
  characteristic collective excitation modes which emerge  as a consequence of the spontaneous violation of the U(1) gauge symmetry; the Higgs mode, the 
  amplitude oscillation of the pair condensate, and 
  the Nambu-Goldstone mode (or the Anderson-Bogoliubov
  mode in neutral superfluid systems), the phase oscillation of the pair condensate.
  In the present study, we explored possible counterpart of the Higgs mode in finite nuclei.
  
  For this purpose, we considered a new kind of pair-transfer operator, the Higgs operator, which probes the
  amplitude motion of the pair condensate. We then described the strength function for the Higgs operator by means of 
  the quasiparticle random phase approximation for the Skyrme-Hartree-Fock-Bogoliubov model. Using the 
  numerical example performed for neutron pairing in Sn isotopes, we have shown that strong
  Higgs response is seen not only in the low-lying pair vibration but also in high-lying pair vibrations
  which have excitation energies up to around 20 MeV. It is more appropriate to deal with strength distribution
  rather than to consider a single pure Higgs mode. We  find also that the calculated Higgs strength function (the one for the Higgs operator) is very close to the incoherent sum of the strength functions defined separately for the pair-addition
  and pair-removal operators, except for the strength of the low-lying pair vibration mode. 
  This indicates that the Higgs response
  can be evaluated  through the pair-addition and pair-removal responses of nuclei.
  
The Higgs response provides us 
  the pair condensation energy, the energy gain caused by the Cooper pair condensate. Considering
an  effective potential curve $U(p)$ as a function of the Cooper pair condensate 
  $p=\langle \Phat_{\mathrm{H}} \rangle$, the Hellmann-Feynman theorem relates the curvature of the potential  
  to  the static polarizability $\alpha_{\mathrm{H}}$ with respect to the Higgs operator, and the latter
  is  directly obtained
  as the inversely energy-weighted sum of the Higgs strength function.
   Furthermore, we have shown using the constrained HFB calculations that the effective
  potential is well approximated by a quartic function as is the case of the Ginzburg-Landau
  phenomenology. Utilizing a simple relation valid for the quartic potential, we arrive at an expression
  of the pair condensation energy, Eq.(\ref{cond-eng}),  which can be evaluated
  in terms of the Higgs polarizability $\alpha_{\mathrm{H}}$, an integrated Higgs response. The validity and accuracy of
  this evaluation is demonstrated using systematic numerical calculations performed for even-even Sn isotopes.
  Possible application of this scheme to the pair transfer experiments is a subject
  to be studied in future.
    
 \section*{Acknowledgement}
 The authors deeply thank S.~Shimoura, S.~Ota, M.~Dozono and K.~Yoshida for stimulating and fruitful
 discussions held in various stages of the present study. This work was supported by the 
JSPS KAKENHI (Grant No. 20K03945).

 \section*{Appendix}
 The matrix elements of the one-body operators  $\hat{\rho}_\alpha(\vecr)=\hat{\rho}(\vecr),\hat{P}(\vecr),\hat{P}^\dagger(\vecr)$
 for two-quasiparticle states  $\ket{ij}$, which appears in the unperturbed response function
 in the spectrum representation Eq.(\ref{resp-fn}),
 are given by
\begin{align}
\braket{ij}{\hat{\rho}_\alpha(\vecr)}{0}& =\sum_\sigma \phi_i^\dagger(\vecrs){\calA}_{\alpha} \bar{\phi}_{\tilde{j}}(\vecrs), \\
\braket{0}{\hat{\rho}_\alpha(\vecr)}{ij}&=\sum_\sigma \bar{\phi}_{\tilde{j}}^\dagger(\vecrs){\calA}_{\alpha} \phi_{i},(\vecrs),
\end{align}
where $\bar{\phi}_{\tilde{j}}(\vecrs)=(-\varphi_{2,j}^*(\vecrst),\varphi_{1,j}^*(\vecrst))^T$. 
The $2 \times 2$ matrix ${\calA}_\alpha$ is 
\beq \label{qp-matel}
{\calA}_\alpha=
\left(
\begin{array}{cc}
 2 & 0 \\
 0 & 0
 \end{array}
 \right)
 ,
 \left(
 \begin{array}{cc}
 0 & 0 \\
 1 & 0
 \end{array}
  \right)
  ,
  \left(
 \begin{array}{cc}
 0 &1 \\
 0 & 0
 \end{array}
  \right)
 \eeq
for  the density  $\hat{\rho}_\alpha(\vecr)=\hat{\rho}(\vecr),\hat{P}(\vecr),\hat{P}^\dagger(\vecr)$, respectively,
which appear in the l.h.s. of Eq.(\ref{rpa}). For the density  $\hat{\rho}_\beta(\vecr)$ which cause
the perturbation,  
the matrix ${\calB}_\beta$ is
${\calB}_{\beta}=
\left(
\begin{array}{cc}
 1 & 0 \\
 0 & -1
 \end{array}
 \right)
 $
 for $\hat{\rho}_\beta(\vecr)=\hat{\rho}(\vecr)$, and the same as Eq.(\ref{qp-matel}) for 
 $\hat{\rho}_\beta(\vecr)=\hat{P}(\vecr)$ and $\hat{P}^\dagger(\vecr)$.

The continuous nature of the unbound quasiparticle states can be taken into account in the
unperturbed response function by using the Green's function for the quasiparticle states:
\begin{align}
 R_{0}^{\alpha\beta}(\vecr,\vecr',\omega) =\frac{1}{4\pi i}
 \int_C dE \sum_{\sigma\sigma'}
& \left\{ {\mathrm Tr}\calA_\alpha \calG_0(\vecrs\vecrsp,E+\hbar\omega+i\epsilon)\calB_\beta 
 \calG_0(\vecrsp\vecrs,E)  \right. \nonumber \\ 
&  +
\left.
{\mathrm Tr}\calA_\alpha \calG_0(\vecrs\vecrsp,E)\calB_\beta 
 \calG_0(\vecrsp\vecrs,E-\hbar\omega-i\epsilon) 
\right\}
\end{align}
where $\calG_0(\vecrsp\vecrs,E)$ is the Green's function for the HFB equation (\ref{HFB}), and
$\int_C dE$ is a contor integral in the complex $E$ plane. Details are given in Ref.\cite{Matsuo2001}.

\end{document}